\documentclass[preprint,3p,twocolumn,12pt]{elsarticle}
\usepackage{amssymb}
\usepackage{amsmath}
\usepackage{slashbox}
\usepackage{color}

\journal{Astroparticle Physics}

\begin{document}

\begin{frontmatter}



\title{Tensor Fluctuations in the Early Universe}


\author{F. Melia}\ead{fmelia@email.arizona.edu}

\address{Department of Physics, The Applied Math Program, and Department of Astronomy, \\
The University of Arizona, AZ 85721, USA}

\begin{abstract}
In standard inflationary cosmology, scalar and tensor perturbations grew
as the Universe expanded and froze when their wavelengths exceeded the Hubble horizon,
producing a tell-tale signature in the fluctuation spectrum and amplitude of the cosmic
microwave background (CMB). But there are now very good reasons to examine whether
structure formation could also have begun via the seeding of quantum fluctuations in
a non-inflationary field. In this {\it Letter}, we study and compare the scalar and
tensor modes produced in these two scenarios, and demonstrate that upcoming observations
to measure the B-mode polarization of the CMB may be able to differentiate between them.
Whereas both scalar and tensor modes should be observable if the field was inflationary,
only scalar modes would be present in the CMB if it were not. Should gravity be purely
classical, however, the tensor modes would have avoided canonical quantization in all
cases, resulting in unmeasurably weak gravitational waves. A non-detection of B-mode
polarization would thus not completely rule out inflation.
\end{abstract}


\end{frontmatter}

% \linenumbers

\section{Introduction}
Quantum fluctuations (QFs) in the early Universe are generally discussed in the context 
of inflation \cite{Guth:1981,Linde:1982}. Indeed, the idea that large-scale structure 
might have started with perturbations in the inflaton field, and the fluctuations they
induced in the spacetime metric, emerged quite early in the development of 
inflationary theory \cite{Mukhanov:1992}.

But as we study the cosmic microwave background (CMB) with higher and higher precision, 
a simple picture in which inflation could have solved the well-known temperature horizon 
problem, while simultaneously seeding and growing the primordial power spectrum, becomes 
increasingly difficult to reconcile with the actual observations. For example, the lack 
of angular correlation at $\theta\gtrsim 60^\circ$ in the CMB contrasts heavily with the 
predictions of standard inflationary cosmology \cite{Guth:1981,Linde:1982,Mukhanov:1992,Copi:2009}, 
though there is still some lingering debate concerning whether instrumental or observational 
selection effects might have caused this deficiency. Even arguments based on cosmic variance 
disfavor the conventional picture at $\gtrsim 3\sigma$. Recent work has instead revealed that 
a more likely reason for this large-scale anomaly is a hard cutoff, $k_{\rm min}=4.34\pm0.50/ 
r_{\rm cmb}$, where $r_{\rm cmb}$ is the comoving distance to the surface of last scattering, 
in the (scalar) fluctuation power spectrum $P_s(k)$ \cite{MeliaLopez:2018,Melia:2021a,Sanchis-Lozano:2022}. 
A zero $k_{\rm min}$ is ruled out by the latest {\it Planck} data \cite{Planck:2018} at over $8\sigma$. 

Generic slow-roll inflation cannot easily accommodate such a spectral cutoff because $k_{\rm min}$ 
signals the first mode to have crossed the Hubble horizon and freeze during the quasi-de Sitter expansion 
\cite{LiuMelia:2020}. It would fix the time at which inflation could have started which, unfortunately,
would have been so late that inflation could then not have accelerated the Universe over the required 
$\gtrsim 60$ e-folds of expansion to eliminate the horizon problem.

As it turns out, however, many of the attractive features associated with the seeding of QFs 
in an incipient scalar field are not at all reliant on an early phase of inflated expansion.  
In fact, over a series of papers, we have begun to promote the idea that a non-inflationary 
scalar field, which we call `numen' to distinguish it from a true inflaton field, actually 
possesses several features distinct from the latter that allow it to satisfy all of the currently 
known restrictions imposed by the CMB observations. For reference, the most significant, relevant 
measurements to date are (i) the scalar spectral index, $n_s=0.9649\pm 0.0042$, in the power 
spectrum $P_s(k)=A_s(k/k_0)^{n_s-1}$ \cite{Planck:2018}, where $k_0=0.05$ Mpc$^{-1}$ is the
pivot scale; (ii) the amplitude $A_s=(2.1\pm 0.04) \times 10^{-9}$ \cite{Planck:2018} of the 
scalar fluctuations; (iii) the aforementioned hard cutoff $k_{\rm min}=4.34\pm0.50/r_{\rm cmb}$ 
\cite{MeliaLopez:2018,Melia:2021a,Sanchis-Lozano:2022}; and (iv) an upper limit to the tensor 
to scalar ratio $r\equiv A_t(1)/A_s(1)\lesssim 0.05$, with $A_t(k/k_0)$ the analog of $A_s(k/k_0)$ for 
tensor modes \cite{Tristram:2020}, which will be a focus of this paper. As we shall see, 
none of these measurements argues {\it against} the influence of a scalar field, $\phi$, nor the 
anisotropies arising from its QFs, $\delta\phi$, but there are now good reasons to question 
whether the potential $V(\phi)$ must be truly inflationary.

In previous work \cite{Melia:2017a,Melia:2019,Melia:2021b}, we demonstrated how scalar fluctuations 
created and amplified by a numen field naturally account for the first three of these observational 
constraints. The primary goal of this {\it Letter} is to address the fourth, given that a focus 
of several upcoming observational campaigns will be to detect and measure the so-called B-mode polarization 
in the CMB (e.g., LiteBIRD \cite{Hazumi:2019}; COrE \cite{Delabrouille:2018}; PRISM \cite{Andre:2014}; 
PICO \cite{Hanany:2019}; and, more recently, ECHO \cite{Adak:2021}). This signal is widely believed 
to be a tell-tale signature of gravity waves produced during inflation, and is thus greatly anticipated 
to provide concrete evidence for the role played by an inflaton field in the early Universe. Actually, 
this is not entirely true if gravity turns out to be purely classical \cite{Ashoorioon:2014}, in which 
case, gravity waves (i.e., tensor modes) produced during inflation would be too weak to detect so,
contrary to the generally held view, the absence of B-mode polarization would not be sufficient to
completely rule out inflation.

Our approach in this {\it Letter} is to assume that tensor modes were indeed quantized at inception, 
regardless of which scalar field may have been present at the time.  In other words, we do not consider 
the possibility that scalar modes began as quantum fluctuations, while tensor modes were purely classical. 
We aim to examine the observable consequences of gravity waves produced by a numen field compared with 
their counterparts created during inflation, complementing our previous comparative study of the scalar 
perturbations. Should gravity waves be detected via the CMB anisotropies, the predicted differences, 
e.g., in the value of $r$, could then be used to meaningfully discern between the inflaton and numen 
scenarios. 

\section{Theoretical Background}
We shall rely on a minimalist set of assumptions, including that the background dynamics is 
dominated by a single homogeneous minimally-coupled scalar field with action
\begin{equation}
S=\int d^4x\sqrt{-g}\;{\mathcal{L}}(\phi,\partial_\mu\phi)\;,\label{eq:S}
\end{equation}
where $\sqrt{-g}=a^3(t)$ for the Friedmann-Lema\^itre-Robertson-Walker (FLRW) metric, 
$a(t)$ is the expansion factor, and the Lagrangian density is given as
\begin{equation}
{\mathcal{L}} = {m_{\rm P}^2\over 16\pi}{\mathcal{R}}+{1\over 2}\partial^\mu\phi
\partial_\mu\phi-V(\phi)\;.\label{eq:L}
\end{equation}
In this expression, $m_{\rm P}\equiv G^{-1/2}$ is the Planck mass, ${\mathcal{R}}$ is the Ricci 
scalar, and $V(\phi)$ is the field potential. As we shall see, although $V(\phi)$ is not known precisely 
for inflation, it is actually specified uniquely for a numen field. We shall derive it shortly.

The foundational principle in the $R_{\rm h}=ct$ universe is that the equation-of-state
in the cosmic fluid corresponds to the zero active mass condition in general relativity,
i.e., $\rho+3p=0$, where $\rho$ and $p$ are the total energy density and pressure,
respectively. This constraint is also inferred empirically from the completed analysis of over 
27 different kinds of observation at low and high redshifts (see Table~2 of
ref.~\cite{Melia:2018a}). Such a universe lacks any horizon problems \cite{Melia:2013,Melia:2018b}, 
so it does not require an early phase of inflated expansion. In keeping with our minimalist
approach in this {\it Letter}, we shall therefore assume that the incipient scalar field
in the $R_{\rm h}=ct$ universe did not inflate.

As long as the numen field satisfies the Cosmological principle, the background contribution,
$\phi_0$, is homogeneous, and therefore the energy density $\rho_\phi$ and pressure $p_\phi$ are 
simply given as
\begin{equation}
\rho_\phi={1\over 2}{\dot{\phi}}^2+V(\phi)\;,
\end{equation}
and
\begin{equation}
p_\phi={1\over 2}{\dot{\phi}}^2-V(\phi)\;.
\end{equation}
With the zero active mass condition, $\rho_\phi+3p_\phi=0$, the potential must therefore
have the form $V(\phi)={{\dot{\phi}}^2}$, whose explicit solution may be written
\begin{equation}
V^{\rm num}(\phi)=V_0\,\exp\left\{-{2\sqrt{4\pi}\over m_{\rm P}}\,\phi\right\}\;.\label{eq:Vphi}
\end{equation}
We therefore recognize the numen field as being a special member of the set of minimally coupled
scalar fields designed to produce so-called power-law inflation
\cite{Abbott:1984,Lucchin:1985,Barrow:1987,Liddle:1989}. But unlike the others, numen is the sole
member of this class that does {\it not} inflate, since $a(t)=t/t_0$, with $t_0$ the present
age of the Universe. 

The potential in Equation~(\ref{eq:Vphi}) is interesting theoretically because it would be 
consistent with Kaluza-Klein cosmologies, string theories and even supergravity. Thus, there already
exists some theoretical motivation to consider the behavior of quantum fluctuations in a numen field, 
even beyond the justification one may derive from the zero active mass condition in general relativity.

We shall assume linearized fluctuations and define the scalar-field and metric perturbations in FLRW 
according to the expressions
\begin{equation}
\phi(t,{\bf x})=\phi_0(t)+\delta\phi(t,{\bf x})\;,\label{eq:phi}
\end{equation}
and
\begin{equation}
g_{\mu\nu}(t,{\bf x})=g_{0\,\mu\nu}(t)+\delta g_{\mu\nu}(t,{\bf x})\;,\label{eq:gmn}
\end{equation}
respectively, where $\phi_0$ and $g_{0\,\mu\nu}$ are the corresponding homogeneous (or background) 
quantities.  Specifically for the FLRW metric, the quantities $\delta g_{\mu\nu}(t,{\bf x})$ are
represented by the various terms in the perturbed line element
\cite{Bardeen:1980,Kodama:1984,Mukhanov:1992,Bassett:2006}
\begin{eqnarray}
\hskip-0.5in ds^2 &=& (1+2A)\,dt^2-2a(t)(\partial_iB)\,dt\,dx^i-\nonumber\\ 
&\null& a^2(t)\left[(1-2\psi)\delta_{ij}+2(\partial_i
\partial_jE)+\right.\nonumber\\
&\null& \left. h_{ij}\right]\,dx^i\,dx^j\,,\quad\label{eq:FLRW}
\end{eqnarray}
where indices $i$ and $j$ denote spatial coordinates, and $A$, $B$, $\psi$ and $E$ describe 
the scalar metric perturbations, while $h_{ij}$ are the tensor perturbations. Throughout 
this {\it Letter}, we work with natural units, in which $\hbar=c=1$.

Various combinations of the perturbed quantities $\delta\phi(t,{\bf x})$ and $\delta g_{\mu\nu}(t,{\bf x})$
produce the scalar and tensor mode distributions, as already shown in great detail in several 
seminal works, including the excellent accounts in refs.~\cite{Bardeen:1980,Kodama:1984,Mukhanov:1992,Bassett:2006}.
But for completeness, we shall here trace a few of the key steps to extract the essential features 
of these scalar and tensor modes, in order to better compare how they are produced in the $R_{\rm h}=ct$
universe versus the standard inflationary scenario, and how their different predictions may be tested
against the observations to identify which model better accounts for the data.

\subsection{Scalar Fluctuations}
The procedure is simplified by identifying in the comoving frame the so-called curvature perturbation 
\cite{Bardeen:1980},
\begin{equation}
\Theta\equiv \psi+\left({H\over \dot{\phi}}\right)\,\delta\phi\;,\label{eq:Theta}
\end{equation}
on hypersurfaces orthogonal to comoving worldlines, as a gauge invariant combination of 
the metric perturbation $\psi$ and the scalar field perturbation $\delta\phi$. We next
expand $\Theta$ in Fourier modes and insert the linearized metric (Eq.~\ref{eq:FLRW}) 
into Einstein's equations, which include the stress-energy tensor written in terms of
$\phi(t,{\bf x})$. This yields the perturbed equation of motion independently for each
mode, $k$, written in the form
\begin{equation}
\Theta_k^{\prime\prime}+2\left({z^\prime\over z}\right)\Theta_k^\prime+k^2\Theta_k=0\;,\label{eq:Thetadyn}
\end{equation}
where prime denotes a derivative with respect to conformal time $d\eta=dt/a(t)$. The parameter
\begin{equation}
z\equiv {a(t)(\rho_\phi+p_\phi)^{1/2}\over H}\label{eq:z}
\end{equation}
plays a pivotal role in this analysis because it is strongly dependent on the chosen background
cosmology. As a result of this dependence, the solutions to Equation~(\ref{eq:Thetadyn})
are quite different for the numen and inflaton fields.

In the conventional approach to solving Equation~(\ref{eq:Thetadyn}), one defines a new
variable, $u_k\equiv z\Theta_k$, known as the Mukhanov-Sasaki variable \cite{Langlois:1994},
with which this expression may be recast in a form (known as the Mukhanov-Sasaki equation)
that permits a more direct physical interpretation:
\begin{equation}
u_k^{\prime\prime}+\left(k^2-{z^{\prime\prime}\over z}\right)u_k=0\;.\label{eq:uk}
\end{equation}
As we shall see shortly, this expression reduces to that of a harmonic oscillator with
a time-dependent frequency for all cosmologies except one. The sole exception is the
$R_{\rm h}=ct$ universe, in which the numen-field fluctuations are described by a 
simple harmonic oscillator equation (analogous to Eq.~\ref{eq:uk}) with a {\it constant} 
frequency.

For the inflaton field, an approximate analytic solution to Equation~(\ref{eq:uk}) 
may be written 
\begin{equation}
u_k^{\rm inf}={e^{-ik\eta}\over\sqrt{2k}}\left(1-{i\over k\eta}\right)\;,\label{eq:ukinf}
\end{equation}
whose initial amplitude, $1/\sqrt{2k}$, is estimated using canonical quantization in the 
so-called Bunch-Davies vacuum \cite{Bunch:1978}. This argument assumes that in the remote 
past ($|\eta|\rightarrow \infty$), where the scale of the fluctuations is much smaller than 
the gravitational horizon \cite{Melia:2018c}, spacetime curvature effects in the field
Hamiltonian may be ignored, so quantum fields behave as they would in Minkowski space
\cite{Bassett:2006}. 

For a numen field, on the other hand, it is straightforward to see that $\eta=t_0\ln a(t)$,
when the zero of conformal time is chosen to coincide with the present age of the Universe, 
and so $z=m_{\rm P}\,a(t)/\sqrt{4\pi}$. In that case, $z^\prime/z=1/t_0$ and
$z^{\prime\prime}/z=1/t_0^2$. The expression analogous to Equation~(\ref{eq:uk}) for
numen is therefore
\begin{equation}
u_k^{\prime\prime}+\alpha_k^2 u_k=0\;,\label{eq:uknum}
\end{equation}
where
\begin{equation}
\alpha_k\equiv {1\over t_0}\sqrt{\left(2\pi R_{\rm h}\over \lambda_k\right)^2-1}\;,\label{eq:alpha}
\end{equation}
and $\lambda_k\equiv 2\pi a(t)/k$ is the proper wavelength of mode $k$. In this
expression, $R_{\rm h}\equiv c/H=ct$ is the apparent/gravitational radius \cite{Melia:2018c},
which defines the Hubble horizon in a spatially flat universe. The principal difference
between Equations~(\ref{eq:uk}) and (\ref{eq:uknum}) is that the numen frequency
$\alpha_k$ is {\it always} independent of time. One can understand this from the fact
that both $R_{\rm h}$ and $\lambda_k$ scale linearly with time, so that $R_{\rm h}/\lambda_k$
(or, equivalently, $kR_{\rm h}/a$) is constant for each mode $k$.

As is well known in inflationary cosmology, the QFs grow much faster than the Hubble horizon
during inflation, and then a reversal occurs after inflation ends. The inflaton modes thus 
cross back and forth across $R_{\rm h}$. On the other hand, given that $R_{\rm h}/\lambda_k$ 
is constant for a scalar field in the $R_{\rm h}=ct$ universe, numen QFs grow 
precisely at the same rate as $R_{\rm h}$ and thus never cross the Hubble horizon. Once 
$\lambda_k$ is established upon exiting (at the Planck scale) into the semi-classical universe,
it remains a fixed fraction of $R_{\rm h}$ forever. Sub-horizon modes in this cosmology
are always sub-horizon, while superhorizon modes are likewise always superhorizon. And
since $\alpha_k$ is independent of time, there is a clear dichotomy in the behavior of the
modes above and below the horizon. Equation~(\ref{eq:uknum}) has simple analytic solutions:
\begin{equation}
u_k^{\rm num}(\eta) = \left\{ \begin{array}{ll}
         B(k)\,e^{\pm i\alpha_k\eta} & \mbox{($2\pi R_{\rm h}>\lambda_k$)} \\
         B(k)\,e^{\pm |\alpha_k|\eta} & \mbox{($2\pi R_{\rm h}<\lambda_k$)}\end{array} 
\right. \;.\label{eq:uknumsol}
\end{equation}
One can see that all sub-horizon numen modes oscillate, while their superhorizon
counterparts do not. At least in this regard, numen modes mirror the dynamic behavior of 
inflaton QFs above and below the horizon.

Aside from the difference in the mode frequency between the inflaton and numen modes,
there is another critical distinction that needs to be emphasized. As noted earlier,
the inflaton mode $u_k^{\rm inf}$ is normalized in a Bunch-Davies vacuum below the 
Planck scale \cite{Bunch:1978}, where wavelengths shorter than the Compton wavelength
$\lambda_{\rm C}$, which coincides with the Planck length, are difficult to interpret 
quantum mechanically.

The numen field does not have this problem. Its QFs would have emerged at the Planck 
scale itself, and the zero active mass condition ensures that the frame into which 
they emerged out of the Planck domain is perfectly geodesic. That is, although the 
Hubble frame has always been expanding in $R_{\rm h}=ct$, it has nevertherless always
been in free fall, with zero internal acceleration. One can easily see this from the
fact that the frequencies $\alpha_k$ in Equation~(\ref{eq:alpha}) are time-independent, 
given that both $R_{\rm h}$ and $\lambda_k$ scale with time in the same way. One may
therefore simply invoke canonical quantization at the Planck scale itself---without
the need to use a Bunch-Davies vacuum---yielding $B(k)=1/\sqrt{2\alpha_k}$ 
for the numen QFs in Equation~(\ref{eq:uknumsol}). 

To clarify these statements, we reiterate that both approaches select a particular
vacuum state at an initial time when the modes are sub-horizon, where the spacetime
in principle resembles a Minkowski vacuum. The principle difference between
them has to do with the possible emergence of a trans-Planckian anomaly in the
case of a Bunch-Davies vacuum, and not for the numen fluctuations exiting into the
semi-classical Universe at the Planck scale itself. The issue here is that the
inflaton fluctuations below the Planck scale have a wavelength much smaller than
the Planck length, which happens to be the Compton wavelength at that energy scale.
It is therefore not clear whether standard quantum mechanics can adequately describe 
the seeding and growth of such modes while they are sub-Planckian 
\cite{Martin:2001,Brandenberger:2013}. 

Before proceeding with the calculation of the scalar power spectrum $P_s(k)$ from Equations~(\ref{eq:ukinf})
and (\ref{eq:uknumsol}), let us briefly digress to explain why one should expect to see a near 
scale-free distribution in both cases, even though the physical mechanism is quite different. 

Cosmologically relevant fluctuations in the inflaton field are created quantum mechanically 
well inside the Hubble radius, where $k\gg aH$. During inflation, however, $H$ varies 
slowly, while the comoving scale $k^{-1}$ remains constant. As such, inflaton fluctuations
exit the horizon and freeze in the super horizon regime, where $k\ll aH$. One may therefore
compute the scalar power spectrum for the inflaton fluctuations at horizon crossing, where
$a(\eta_\oplus)H(\eta_\oplus)=k$. As we shall see from the definition $u_k=z\Theta_k$, this
progression towards higher values of $a$ with increasing $k$ produces the near scale-free
$P_s^{\rm inf}(k)$. It would have been perfectly scale free if $H$ were strictly constant during
inflation. The slight difference between $n_s=0.9649$ and $1$ is attributed to the fact
that $H$ varies slowly due to the presumed slow-roll inflaton potential, $V^{\rm inf}(\phi)$. 

For the numen field, on the other hand, there is never any horizon crossing. Instead, each
mode $k$ emerges into the semi-classical Universe when its wavelength $\lambda_k$ equals
$2\pi\lambda_{\rm P}$, where $\lambda_{\rm P}\equiv \sqrt{4\pi G}$ is the Planck wavelength,
after which the sub-horizon modes continue to evolve according to the oscillatory solution in 
Equation~(\ref{eq:uknumsol}). Thus, each succeeding $k$ emerges at progressively later times 
(and hence larger values of $a$), creating an analogously near scale-free scalar power spectrum 
$P_s^{\rm num}(k)$. The scalar index $n_s$ is slightly less than $1$ in this case because
the so-called Hubble friction term in Equation~(\ref{eq:Thetadyn}) produces a small difference 
between $\alpha_k$ and $k$ in Equation~(\ref{eq:alpha}) (see also Eq.~\ref{eq:Psnumfinal} below).

Incidentally, we should point out at this stage that one of the strongest arguments in favor of the
superhorizon freeze-out mechanism during inflation also applies to the emergence of the numen field 
fluctuations out of the Planck regime. The observed CMB fluctuations are coherent, which requires
all of the Fourier modes with the same $k$ to have an identical phase \cite{Dodelson:2003}. This 
happens for the inflaton field because all the modes with the same $k$ cross the horizon at the 
same time. For the numen QFs, all the modes with the same $k$ also have the same phase because
they emerge out of the Planck domain at the same time $t_k$, defined by the condition
$\lambda_k=\lambda_{\rm P}$, where $\lambda_k=2\pi a(t_k)/k$. In fact, the numen mechanism 
is simpler than its inflaton counterpart because it requires fewer steps and avoids some of the 
difficulties faced by conventional inflation.

Finally, the scalar power spectrum is defined from the coefficient of the Fourier transform of the
two-point correlation function,
\begin{eqnarray}
\hskip-0.2in \langle \Theta(x)\Theta(y)\rangle&\hskip-0.1in=\hskip-0.1in&\int{d^3k\over 
(2\pi)^3}{d^3k^\prime\over(2\pi^3)}\langle
\Theta_k\Theta_{k^\prime}\rangle\times\nonumber\\
&\null& e^{i{\bf k}\cdot{\bf x}} e^{i{\bf k^\prime}\cdot{\bf y}}\;,\label{eq:twopoint}
\end{eqnarray}
where $\langle ...\rangle$ denotes an ensemble average of the fluctuations. Specifically, we define
$P_\Theta(k)$ from the expression
\begin{equation}
\langle\Theta_k\Theta_{k^\prime}\rangle=(2\pi)^3\delta({\bf k}+{\bf k^\prime})P_\Theta(k)\;,\label{eq:PTheta}
\end{equation}
from which one may infer the dimensionless power spectrum 
\begin{equation}
P_s(k)\equiv {k^3\over 2\pi^2}P_\Theta(k)\;.\label{eq:Ps}
\end{equation}

The inflaton scalar power spectrum $P_s^{\rm inf}(k)$ then follows from the definition of
$u_k^{\rm inf}$ and Equations~(\ref{eq:z}), (\ref{eq:ukinf}) and (\ref{eq:Ps}), taken in
the superhorizon limit, where $|k\eta|\ll 1$:
\begin{equation}
P_s^{\rm inf}(k)={H(\eta_\oplus)^2\over(2\pi)^2}{H(\eta_\oplus)^2\over \dot{\phi}^2_\oplus}\;.\label{eq:Psinf}
\end{equation}

The analogous calculation for numen fluctuations, using the definition $u_k^{\rm num}=a\Theta_k/\sqrt{4\pi G}$
and Equations~(\ref{eq:alpha}), (\ref{eq:uknum}) and (\ref{eq:Ps}), gives
\begin{equation}
P_s^{\rm num}(k)={G\over\pi}\left({k\over a}\right)^2\left[1-\left({k_{\rm min}^{\rm num}\over 
k}\right)^2\right]^{-1/2}\;,\label{eq:Psnum}
\end{equation}
where $k_{\rm min}^{\rm num}\equiv 1/t_0$ constitutes a cutoff in the numen scalar power spectrum. Physically,
this wavenumber corresponds to the very first mode that could have emerged out of the Planck regime
at about the Planck time following the Big Bang \cite{Melia:2019}. It appears to be fully consistent 
with the cutoff $k_{\rm min}$ measured in the angular correlation function of the CMB \cite{MeliaLopez:2018}.

But notice that the numen power spectrum, $P_s^{\rm num}(k)$, is not yet on an equal footing with its 
inflaton counterpart, $P_s^{\rm inf}(k)$. The reason is that in order to determine when the QFs
stop oscillating and devolve into features growing classically under their own self-gravity, one
must introduce some new physics. In the case of inflation, one assumes that the QFs cease oscillating 
once they cross the horizon at time $\eta_\oplus$. $P_s^{\rm inf}(k)$ is 
therefore fully dependent on this crossing and the viability of this assumption. Numen 
fluctuations never cross the horizon, however, so this argument does not apply to them.  
The power spectrum, $P_s^{\rm num}(k)$, in Equation~(\ref{eq:Psnum}) is thus still a function 
of $a$. A new mechanism must be introduced to end the oscillations, which we 
shall discuss next.

\subsection{The Quantum to Classical Transition}
Once the QFs cross the horizon in standard inflationary cosmology, the assumption is made 
that they classicalize and freeze, though there is no complete theoretical 
understanding of how this process actually takes place 
\cite{Penrose:2004,Perez:2006,Mukhanov:2005,Weinberg:2008,Lyth:2009,Bengochea:2015,Melia:2021b}. 
If this mechanism does in fact work, the fluctuations then begin to grow classically under 
their own self-gravity upon re-entering the horizon after inflation has ended. The amplitude, 
$A_s^{\rm inf}$, inferred from Equation~(\ref{eq:Psinf}), is therefore set by the overall 
evolution of the modes in Equation~(\ref{eq:ukinf}) between the time they were seeded in 
the Bunch-Davies vacuum and the time ($\eta_\oplus$) at which they first crossed the horizon.

Irrespective of whether classicalization can occur in this fashion, however, one cannot 
invoke a horizon crossing to terminate the quantum oscillation of numen modes. But regardless 
of which scalar field one adopts, it must eventually decay into standard model particles in
order to (re)initiate the hot Big Bang expansion of the Universe. In the case of the inflaton
field, this transition is assumed to take place at the end of inflation via as-yet-unknown
extensions to the standard model of particle physics. In an analogous fashion, the numen
quantum oscillations will cease---and a classical growth of the fluctuations begins---once 
the numen field similarly devolves into standard model particles. The time at which this
transition occurs is constrained by the amplitude $A_s(k)$ measured in the scalar power
spectrum.

The possible transition of sub-horizon modes from quantum oscillators to classically
growing fluctuations has already been discussed extensively in the literature. For
example, during (re)heating \cite{Kofman:1997,Allahverdi:2010,Berera:2020} the scalar 
field previously driving the cosmic expansion decays into other particles that then
thermalize via collisions. The scalar field itself need not be inflationary.
Indeed, the particle production on its own probably leads to decoherence 
\cite{Calzetta:1990,Son:1996,Shtanov:1995}.

Quite generically, then, one may describe the dynamics of a transition from quantum 
oscillations to classically growing fluctuations in terms of a particular length 
(or energy) scale, that we shall call $L_*$, analogous to the Planck scale
$\lambda_{\rm P}$ \cite{Brouzakis:2012,Dvali:2012}. As we shall see below, 
this transition scale---when combined with the observed fluctuation spectrum in
the CMB---serves the principal purpose of establishing the amplitude $A_s^{\rm num}(k)$. 

It is important to emphasize that we view {\it both} the transition to
standad model particles and the classicalization of the numen quantum fluctuations
to have occurred at the same energy scale. Since the numen decay process may itself 
lead to decoherence, the quantum fluctuations may have classicalized during this
decay. 

Nevertheless, there is still a great deal of uncertainty concerning
the nature and scale associated with $L_*$. We know that quantum fluctuations must 
have classicalized for them to create the large-scale structure we see, but we do
not yet know how or where that happened. Here, we invoke the idea that they 
classicalized when the incipient scalar field devolved into standard model particles. 
As noted above, this process has been discussed previously, e.g., by 
refs.~\cite{Calzetta:1990,Son:1996,Shtanov:1995}, who argued that the particle 
production on its own can lead to decoherence, which may be the mechanism by which 
the quantum fluctuations acquire classical features. In this regard, the scale $L_*$ 
should not be the Planck scale, where the energy is orders of magnitude greater than 
the energy scale in grand unified theories (GUTs). As we shall see shortly, what is 
particularly interesting about this concept is that the value of $L_*$ required
by the observed amplitude of the fluctuations in the CMB is in fact very close to 
the GUT scale, which provides some evidence in support of the view that the
numen scalar field may have transitioned to standard model particles, given that
the GUT scale could arguably be associated with physics underlying this process. 

In this context, the quantum oscillation of mode $k$ will cease at 
$a(t_k^*)=L_*f(k)k/2\pi$, corresponding to cosmic time $t_k^*=t_0L_*f(k)k$. Since
we do not yet understand the exact nature of the classicalization scale, we allow
for a possible---presumably weak---dependence of this transition on $k$ itself,
which we represent with $f(k)$. Given the primitive state of this model, the
inclusion of such detail would seem to be premature. But the spectral index
has been measured to very high precision, and we shall gauge the importance
of $f(k)$ via its impact on the predicted value of $n_s$. 

At this point, the numen scalar power spectrum freezes out, and 
Equation~(\ref{eq:Psnum}) takes its final form, 
\begin{equation}
P_s^{\rm num}(k)=\left({\lambda_{\rm P}\over 2\pi L_*f}\right)^2\left[1-
\left({k_{\rm min}^{\rm num}\over k}\right)^2\right]^{-1/2}\hskip-0.2in.\label{eq:Psnumfinal}
\end{equation}
When compared with the power spectrum observed in the CMB \cite{Planck:2018}, this 
expression provides us with a very revealing interpretation. First, the expression
\begin{eqnarray}
\hskip-0.4in n_s&\equiv&1+{d\ln P_s^{\rm num}\over d\ln k}\nonumber\\
&=&1-2k{f^\prime\over f}-\left[\left({k\over k_{\rm min}^{\rm num}}\right)^2-1\right]^{-1}
\quad \label{eq:ns}
\end{eqnarray}
suggests that a good match to the observed scalar index in the sample $k$-range
may be obtained with 
\begin{equation}
f(k)=\left({k\over k_{\rm min}^{\rm num}}\right)^{0.018}\;.\label{eq:fk}
\end{equation}
The actual scale for $k$ on the right-hand side is arbitrary, but we write it
this way to reduce the number of unknowns. This choice of $f(k)$ also provides
a reasonable match with the `running of the spectral index,' $\alpha_s\equiv
dn_s/d\ln k$, whose measured value is $\alpha_s=-0.0045\pm 0.0067$ \cite{Planck:2018}.
As expected, the dependence of the classicalization scale on $k$ is very weak.

Second, one finds from the measured value of $A_s$ (at the pivot wavenumber
$k_0=0.05$ Mpc$^{-1}$) that $L_*$$\sim$$3.5\times 10^3 \lambda_{\rm P}$. 
Thus, given a Planck scale $m_{\rm P}\approx 1.22\times 10^{19}$ GeV, we conclude 
that the numen field must have devolved into standard model particles at roughly 
$3.5\times 10^{15}$ GeV---remarkably consistent with the energy scale in GUTs. 
We interpret this to mean that the numen QFs oscillated 
until $t\sim 3.5\times 10^3t_{\rm P}$, after which the numen field (including
its fluctuations) devolved into GUT particles and the power spectrum $P_s^{\rm num}(k)$ 
remained frozen thereafter.

As we shall see in the next subsection, this situation is different
from that of the tensor fluctuation spectrum, which instead continued to decay
as $\sim a^{-2}$. At the classicalization scale, the curvature fluctuations
$\Theta_k$ transitioned into classical perturbations of the energy density
$\rho$. The subsequent evolution of the dimensionless fluctuations
$\delta_k\equiv \delta\rho_k/\rho$ has already been studied and reported
in ref.~\cite{Melia:2017b}. The dynamical evolution of $\delta_k$ is
given by Equation~(27) in that paper when $\rho$ is dominated by matter.
A similar expression may be written for the earlier radiation-dominated
expansion. The evolution of $\rho$ and $\delta\rho$ is quite different
when the equation-of-state corresponds to the zero active mass condition
in general relativity, i.e., $\rho+3p=0$, compared to that in the
standard model. In this scenario, $\rho\sim a^{-2}$ throughout cosmic
history, and $\delta_k$ is only weakly dependent on time. In fact, 
these density fluctuations grew as the Universe expanded, providing a
reasonably good match to the observed `growth factor' and volume-deliminted
variance $\sigma_8$ (see fig.~1 in ref.~\cite{Melia:2017b}).

\subsection{Tensor Fluctuations}
Unlike scalar modes, the tensor fluctuations $h_{ij}$ in Equation~(\ref{eq:FLRW})
are purely gravitational and have no sources. Their dynamics is governed by the
action
\begin{equation}
S={m_{\rm P}^2\over 64\pi}\int d^4x\,\sqrt{-g}\left[h_{ij}^\prime h^{\prime\,ij}-\partial_i h_{jk}
\partial^{\,i} h^{jk}\right],\label{eq:Stensor}
\end{equation}
which is very similar to that of a massless scalar field up to an overall multiplicative
factor. Following a procedure analogous to that for $\Theta$ in Equation~(\ref{eq:Theta}),
we then expand $h_{ij}$ in Fourier modes,
\begin{equation}
h_{ij}=\sum_{q=+,\times}\int {d^3k\over (2\pi)^3}\epsilon^q_{ij}(k)\,h^q_k\,e^{i{\bf k}\cdot
{\bf x}}\;,\label{eq:hijFourier}
\end{equation}
where $q$ here represents the two polarization states, such that $\epsilon^q_{ij}(k)
\epsilon^{q^\prime}_{ij}(k)=2\delta_{q{q^\prime}}$ and $\epsilon_{ii}=k^j\epsilon_{ij}=0$.
The metric fluctuation $h_{ij}$ is symmetric, traceless and transverse, i.e., $h^i_{\; i}=0$
and $\partial_i h^i_{\; j}=0$.

With this Fourier decomposition, the action for the tensor fluctuations becomes
\begin{equation}
S=\sum_q{m^2_{\rm P}\over 32\pi}\int d\eta\,d^3k\, a^2\left[(h^{\prime\,q}_k)^2-k^2(h^q_k)^2\right].
\end{equation}
And if we next introduce a second Mukhanov-Sasaki variable, 
\begin{equation}
v^q_k\equiv {m_{\rm P}a\over\sqrt{32\pi}}h^q_k\;,\label{eq:v}
\end{equation}
it is not difficult to show that each polarization mode must satisfy the equation
\begin{equation}
(v^q_k)^{\prime\prime}+\left(k^2-{a^{\prime\prime}\over a}\right)v_k^q=0\;,\label{eq:vqk}
\end{equation} 
complementary to Equation~(\ref{eq:uk}) for the scalar modes.

We must now decide whether or not to treat these tensor modes quantum mechanically. 
As noted earlier, gravity may be purely classical \cite{Ashoorioon:2014}, in which 
case the solutions to Equation~(\ref{eq:vqk}) would have an unconstrained amplitude, 
possibly even zero. In this paper, we shall ignore that possibility, and instead
assume that gravity waves generated as metric perturbations in the early Universe 
do indeed begin as quantum fluctuations, described as harmonic oscillators according 
to Equation~(\ref{eq:vqk}), with an initial amplitude set by canonical quantization.

In the case of the inflaton field, each polarization of the tensor mode thus behaves 
as a massless scalar field in de Sitter space, with 
\begin{equation}
h^q_k={\sqrt{32\pi}\over m_{\rm P}a}v^q_k\;,\label{eq:hqk}
\end{equation}
where $v^q_k$ is the solution to Equation~(\ref{eq:vqk}), with a time-varying frequency
and an initial amplitude $1/\sqrt{2k}$. The power spectrum for each polarization mode
is thus extracted from an expression analogous to Equation~(\ref{eq:PTheta}),
\begin{eqnarray}
\hskip-0.3in\langle h^q_k(\eta) h^2_{k^\prime}(\eta)\rangle&\hskip-0.1in=\hskip-0.1in&
(2\pi)^3\delta({\bf k}+{\bf k^\prime}) {32\pi\over m^2_{\rm P}a^2}\times\nonumber\\
&\null&{1\over 2\eta^2 k^3}(1+\eta^2k^2)\;.\label{eq:hq}
\end{eqnarray}

A second, even more critical, assumption must now be made for which, admittedly, there 
is yet no fundamental justification or experimental evidence. In the conventional 
inflationary picture, one assumes that tensor modes `classicalize' as they cross the 
Hubble horizon, mirroring their scalar counterparts. In essence, this fixes the time 
$\eta_\oplus$ at which the tensor power spectrum is to be calculated, 
$a(\eta_\oplus)H(\eta_\oplus)=k$, under the assumption that it remains frozen 
thereafter. And so we infer from this sequence of arguments the dimensionless power 
spectrum of the inflaton tensor fluctuations, taking into account both polarizations 
(i.e., a factor $2$):
\begin{equation}
P_t^{\rm inf}(k)={16H(\eta_\oplus)^2\over \pi m^2_{\rm P}}\label{eq:Ptinf}
\end{equation}
where, as usual, $|k\eta|\ll 1$ in the superhorizon limit.

One can immediately see from Equations~(\ref{eq:Psinf}) and (\ref{eq:Ptinf})
why the inflaton power spectra are almost scale free, though not exactly. The
expansion during slow-roll inflation is characterized by a slowly varying
Hubble parameter, in this case $H(\eta_\oplus)$, which presumably accounts
for the difference between the observed $n_s=0.9649$ and $1$. 

Insofar as the upcoming B-mode polarization measurements of the CMB are concerned, 
one of the most important predictions of this scenario is the tensor power relative 
to that in the scalar modes, defined as
\begin{equation}
r^{\rm inf}\equiv {P_t^{\rm inf}\over P_s^{\rm inf}}\;.\label{eq:rinf}
\end{equation}
It is easy to see from the Friedmann equation that
\begin{equation}
{\dot\phi}^2={3\over 8\pi G}H^2(1+w)\;,\label{eq:dotphi}
\end{equation}
where we have defined the equation of state parameter, $w$, via the expression
\begin{equation}
p_\phi=w\rho_\phi\;.
\end{equation}
Therefore,
\begin{equation}
r^{\rm inf}=16\epsilon\;,
\end{equation}
written in terms of the so-called slow-roll parameter
\begin{equation}
\epsilon\equiv {3\over 2}(1+w)\;.\label{eq:epsilon}
\end{equation}
If inflation were purely de Sitter, we would have $w=-1$ and $r^{\rm inf}$ would be strictly $0$.
Any slow roll at all produces a non-zero $r^{\rm inf}$, but presumably always with $r^{\rm inf}\ll 1$.
As noted in the introduction, the current observational limit appears to be $r\lesssim 0.05$
\cite{Planck:2018}.

Let us now compare this situation with that pertaining to the numen tensor modes. 
In this case, the mode equation reduces to that of a true harmonic oscillator,
\begin{equation}
(v^q_k)^{\prime\prime}+\alpha_k^2v_k^q=0\;,\label{eq:vqknum}
\end{equation}
with a constant frequency, $\alpha_k$, identical to that in Equation~(\ref{eq:alpha}).
Each sub-horizon polarization mode therefore evolves according to the expression 
\begin{equation}
(h^q_k)^{\rm num}={\sqrt{32\pi}\over m_{\rm P}a}{1\over\sqrt{2\alpha_k}}e^{-i\alpha_k\eta}\;.\label{eq:hqkfinal}
\end{equation}

With the definition in Equation~(\ref{eq:hqk}), it is straightforward to see that the combined 
power for the numen tensor modes is thus
\begin{equation}
P_t^{\rm num}(k)={16 G\over\pi}\left({k\over a}\right)^2\left[1-\left({k_{\rm min}^{\rm num}\over 
k}\right)^2\right]^{-1/2}\hskip-0.2in.\label{eq:Ptnum}
\end{equation}
But once generated, these gravitational fluctuations do not couple at all to the matter
(at least to first order), and evolve completely independently of $\delta\phi$ and the
other metric perturbations in Equation~(\ref{eq:FLRW}). The background content of the
Universe, before or after the numen $\phi$ devolves into standard model particles, affects
only the time dependence of $P_t^{\rm num}(k)$ via the expansion factor $a(\eta)$. Therefore,
the mechanism that ends the quantum oscillations of the numen scalar modes, i.e., the 
hypothesized field decay at the GUT scale, has no impact on the sub-horizon tensor modes, which 
presumably continue to evolve according to Equation~(\ref{eq:hqkfinal}). As such, the ratio
\begin{equation}
r^{\rm num}\equiv {P_t^{\rm num}\over P_s^{\rm num}}\;,\label{eq:rnum}
\end{equation}
calculated from Equations~(\ref{eq:Psnumfinal}) and (\ref{eq:Ptnum}), has the form 
\begin{equation}
r^{\rm num}={16\over A_s}\left({\lambda_{\rm P}\over \lambda_k}\right)^2\;,\label{eq:rnumlambda}
\end{equation}
in the terms of the previously defined mode wavelength $\lambda_k\equiv 2\pi a/k$.

A pertinent question concerns when this ratio ought to be measured, since
it evolves in time. Distances in the $R_{\rm h}=ct$ universe grow linearly with $t$, 
so Equation~(\ref{eq:rnumlambda}) may be written in the even simpler form
\begin{equation}
r^{\rm num}={16\over A_s}\left({t_{\rm P}\over t}\right)^2\;,
\end{equation}
where $t$ is the cosmic time at which the fluctuations are {\it observed}. By convention, 
this ratio is `measured' at the pivot scale $k_0=0.05$ Mpc$^{-1}$, so one should choose 
this specific wavenumber for its evaluation, though it turns out not to matter in this 
particular case. Clearly, given that $t\gg t_{\rm P}$ throughout the visible Universe, 
this $r^{\rm num}$ is completely negligible and unobservable using any conceivable 
measurement, including the B-mode polarization of the CMB, for which $t_{\rm CMB}\gg t_{\rm P}$.

\section{Discussion and Conclusion}
We begin by pointing out the fact that the manner in which $n_s$,
$\alpha_s$, $r$ and $A_s$ arise is certainly dependent on the choice of
cosmological model, particularly in terms of whether or not an inflationary
expansion took place in the early Universe. But their values are measured from
the pattern of angular anisotropies in the CMB, independently of the distance 
to the surface of last scattering. As such, it is reasonable to expect that 
these `measured' quantities ought to be largely model-independent. In other
words, the optimization of these parameters in the context of the {\it Planck} 
concordance model should also be relevant to a direct comparison with the
predictions of $R_{\rm h}=ct$.

As we have seen, the equation of motion for gravitational waves seeded in the early Universe
is similar to that of a massless scalar field, except that it does not couple the wave 
amplitude to the matter. As is well known, gravitational waves effectively pass unaffected 
through everything, at least to first order. Thus, numen tensor modes could not have
transitioned into classical gravitational waves when the scalar field devolved into standard
model particles around the GUT scale. Moreover, since sub-horizon tensor modes never 
cross the horizon in a non-inflationary environment, even the mechanism for classicalization 
invoked in standard cosmology---whether physically viable or not---does not apply to them. 
Thus, short of some exotic `collapse' mechanism \cite{Das:2014}, there does not appear to 
be a physical process that would have transitioned quantum tensor modes into classical waves 
early enough to meaningfully impact the CMB.

Our principal result is therefore quite straightforward. If primordial gravity waves were 
seeded as quantum fluctuations, both scalar and tensor modes should be present in the CMB
if the early Universe was dominated by an inflationary scalar field. If this field was
not inflationary, however, the impact of primordial gravitational waves on the CMB B-mode
polarization should be negligibly small and unobservable.

There is a caveat to this conclusion, however. Gravitational waves could also have been 
produced by secondary processes, such as from the interaction of classical matter waves
generated during (re)heating \cite{Khlebnikov:1997} which, in the non-inflationary case, 
would have occurred at the GUT scale. But the induced gravitational-wave power spectrum
of such a process is not scale invariant. It has a pronounced peak whose frequency depends
on the energy scale at which these modes are produced. For example, if the scalar-field
decay occurred at $\sim 10^{15}$ GeV, the peak of the spectrum would lie at $\sim 10^8$ Hz.

Advanced simulations of gravitational-wave creation during (re)heating have also been
carried out for a broad range of environments and assumptions (see, e.g.,
\cite{Easther:2007,Garcia-Bellido:2008,Dufaux:2007,Price:2008,Kusenko:2008}), sometimes 
showing that the amplitude of the peak can be orders of magnitude larger than the 
scale-invariant background of tensor modes produced during the numen or inflaton 
expansion. But the spectrum of these secondary waves is always distinct from that 
of the latter, and easily distinguishable. 

The distinction between the inflaton and numen tensor modes, and the different
impact they would have on B-mode polarization in the CMB, adds an important dimension 
to the probative power of primordial fluctuations in the inflationary and 
non-inflationary scenarios. It is worth highlighting again that the scalar modes 
already provide an important difference in the physical interpretation of the cutoff 
$k_{\rm min}$ \cite{Melia:2019}. In the inflationary context, this minimum wavenumber
signals the time at which inflation could have started, which creates some tension
between the mechanism for producing the observed CMB anisotropies and the number of 
inflationary e-folds required to solve the temperature horizon problem \cite{LiuMelia:2020}.

In the numen context, on the other hand, the observed $k_{\rm min}$ presumably
corresponds to $k_{\rm min}^{\rm num}$, the first mode to have exited the Planck 
regime at about the Planck time. Aside from the fact that this appears to represent
a more `natural' scale associated with known physics, it also offers the possibility
of resolving several long-standing problems with the inflationary paradigm, such as
the so-called trans-Planckian anomaly \cite{Martin:2001,Brandenberger:2013}.

The B-mode polarization in the CMB is much more difficult to measure, for various well-documented
reasons, including the complexity of separating the true signal from the Galactic dust background 
and the fact that $r\ll 1$. But several upcoming experiments, including those mentioned in
the introduction, show much promise in overcoming these limitations. The hope is that at least 
one of them will produce a reliable estimate of the tensor to scalar ratio, beyond merely an upper 
limit. Such a measurement would then overwhelmingly favor an inflationary scenario over the
alternative numen expansion which, as noted earlier, is not expected to produce an observable
primordial tensor signal. 

On the other hand, a new upper limit, even lower than the current value $r\lesssim 0.05$, would 
leave the door wide open---either because the initial scalar field was non-inflationary, or the energy 
scale of inflation was much lower than expected, or $H$ was strictly constant during inflation. 
The tensor to scalar ratio might also be unmeasurable if gravity were purely classical, as unlikely
as that would seem right now, so that primordial gravity waves did not begin as quantum fluctuations
constrained via canonical quantization. The upcoming search for B-mode polarization in the
CMB thus has much to offer. 
 
\section*{Acknowledgements} 
I am grateful to Amherst College for its support through a John Woodruff 
Simpson Lectureship. I am especially grateful to the anonymous
referee for an exceptional review. Their comments and suggestions have led
to significant improvements in the presentation of the material in this
manuscript.



%% References

\bibliographystyle{elsarticle-num}

\begin{thebibliography}{}
\bibitem{Guth:1981} A. H. Guth, Phys. Rev. D {\bf 23} (1981) 347.
\bibitem{Linde:1982} A. Linde, Phys. Lett. B {\bf 108} (1982) 389.
\bibitem{Mukhanov:1992} V. F. Mukhanov, H. A. Feldman and R. H. Brandenberger, Phys. Rep. {\bf 215} (1992) 203.
\bibitem{Copi:2009} C. J. Copi, D. Huterer, D. J. Schwarz and G. D. Starkman, MNRAS {\bf 399} (2009) 295.
\bibitem{MeliaLopez:2018} F. Melia and M. L\'opez-Corredoira, A\&A {\bf 618} (2018) A87.
\bibitem{Melia:2021a} F. Melia, Q. Ma, J.-J. Wei \& B. Yu, A\&A {\bf 655} (2021) A70.
\bibitem{Sanchis-Lozano:2022} M.-A. Sanchis-Lozano, F. Melia, M. L\'opez-Corredoira \& N. Sanchis-Gual,
A\& {\bf 660} (2022) A121.
\bibitem{Planck:2018} Planck Collaboration, A\&A in press (eprint arXiv:1807.06209) (2018).
\bibitem{LiuMelia:2020} J. Liu and F. Melia, Proc. R. Soc. A {\bf 476} (2020) 20200364.
\bibitem{Tristram:2020} M. Tristram et al., A\&A {\bf 647} (2020) A128.
\bibitem{Melia:2017a} F. Melia, CQG {\bf 34} (2017) 015011.
\bibitem{Melia:2019} F. Melia, EPJ-C Lett. {\bf 79} (2019) 455.
\bibitem{Melia:2021b} F. Melia, PLB {\bf 818} (2021) 136632.
\bibitem{Hazumi:2019} M. Hazumi et al., J. Low Temp. Phys. {\bf 194} (2019) 443.
\bibitem{Delabrouille:2018} J. Delabrouille et al., JCAP {\bf 2018} (2018) 014.
\bibitem{Andre:2014} P. Andr\'e et al., JCAP {\bf 2014} (2014) 006.
\bibitem{Hanany:2019} S. Hanany et al., eprint arXiv:1902.10541 (2019).
\bibitem{Adak:2021} D. Adak et al., eprint arXiv:2110.12362 (2021).
\bibitem{Ashoorioon:2014} A. Ashoorioon et al., MPLA {\bf 29} (2014) 1450163.
\bibitem{Melia:2018a} F. Melia, MNRAS {\bf 481} (2018) 4855.
\bibitem{Melia:2013} F. Melia, A\&A {\bf 553} (2013) id. A76.
\bibitem{Melia:2018b} F. Melia, EPJ-C Lett. {\bf 78} (2018) 739.
\bibitem{Abbott:1984} L. F. Abbott and M. B. Wise, Nucl. Phys. {\bf 244} (1984) 541.
\bibitem{Lucchin:1985} F. Lucchin and S. Materrese, PRD {\bf 32} (1985) 1316.
\bibitem{Barrow:1987} J. Barrow, PLB {\bf 187} (1987) 12.
\bibitem{Liddle:1989} A. R. Liddle, PLB {\bf 220} (1989) 502.
\bibitem{Bardeen:1980} J. M. Bardeen, PRD {\bf 22} (1980) 1882.
\bibitem{Kodama:1984} H. Kodama and M. Sasaki, Prog. Theor. Phys. Suppl. {\bf 78} (1984) 1.
\bibitem{Bassett:2006} B. A. Bassett, S. Tsujikawa and D. Wands, Rev. Mod. Phys. {\bf 78} (2006) 537.
\bibitem{Langlois:1994} D. Langlois, CQG {\bf 11} (1994) 389.
\bibitem{Bunch:1978} T. S. Bunch and P.C.W. Davies, Proc. R. Soc. A {\bf 360} (1978) 117.
\bibitem{Melia:2018c} F. Melia, Am. J. Phys. {\bf 86} (2018) 585.
\bibitem{Martin:2001} J. Martin \& R. H. Brandenberger, PRD {\bf 63} (2001) 123501.
\bibitem{Brandenberger:2013} R. H. Brandenberger \& J. Martin, CQG {\bf 30} (2013) 113001.
\bibitem{Dodelson:2003} S. Dodelson, Aip Conf. Proc. {\bf 689} (2003) 184.
\bibitem{Penrose:2004} R. Penrose, {\it The Road to Reality}. Vintage Books, USA (2004) 861.
\bibitem{Perez:2006} A. Perez, H. Sahlmann and D. Sudarsky, CQG {\bf 23} (2006) 2317.
\bibitem{Mukhanov:2005} V. F. Mukhanov, {\it Physical Foundations of Cosmology} Cambridge University Press, UK (2005) 340.
\bibitem{Weinberg:2008} S. Weinberg, {\it Cosmology}. Oxford University Press, USA (2008) 447.
\bibitem{Lyth:2009} D. H. Lyth and A. R. Liddle, {\it The Primordial Density Perturbation} Cambridge University Press, UK (2009) 386.
\bibitem{Bengochea:2015} G. R. Bengochea, P. Canate and D. Sudarsky, PLB {\bf 743} (2015) 484.
\bibitem{Kofman:1997} L. Kofman, A. D. Linde \& A. A. Starobinsky, PRD {\bf 56} (1997) 3258.
\bibitem{Allahverdi:2010} R. Allahverdi et al., ARNPS {\bf 60} (2010) 27.
\bibitem{Berera:2020} A. Berera, S. Brahma \& J. R. Calderon, JHEP {\bf 2020} (2020) 71.
\bibitem{Calzetta:1990} E. Calzetta \& F. D. Mazzitelli, PRD {\bf 42} (1990) 4066.
\bibitem{Son:1996} D. T. Son, e-print hep-ph/9601377 (1996).
\bibitem{Shtanov:1995} Y. Shtanov, J. H. Traschen \& R. H. Brandenberger, PRD {\bf 51} (1995) 5438.
\bibitem{Brouzakis:2012} N. Brouzakis, J. Rizos and N. Tetradis, PLB {\bf 708} (2012) 170.
\bibitem{Dvali:2012} G. Dvali and C. Gomez, JCAP {\bf 1207} (2012) 015.
\bibitem{Melia:2017b} F. Melia, MNRAS {\bf 464} (2017) 1966.
\bibitem{Das:2014} S. Das, S. Sahu, S. Banerjee \& T. P. Singh, PRD {\bf 90} (2014) 043503.
\bibitem{Khlebnikov:1997} S. Y. Khlebnikov \& I. I. Tkachev, PRD {\bf 56} (1997) 653.
\bibitem{Easther:2007} R. Easther \& E. A. Lim, JCAP {\bf 0604} (2007) 010.
\bibitem{Garcia-Bellido:2008} J. Garcia-Bellido \& D. G. Figueroa, PRL {\bf 98} (2008) 061302.
\bibitem{Dufaux:2007} J. F. Dufaux et al., PRD {\bf 76} (2007) 123517.
\bibitem{Price:2008} L. R. Price \& X. Siemens, PRD {\bf 78} (2008) 063541.
\bibitem{Kusenko:2008} A. Kusenko \& A. Mazumdar, PRL {\bf 101} (2008) 211301.
\end{thebibliography}

\end{document}